%% file: main.tex
\documentclass[manuscript,screen]{acmart}

\usepackage{braille}
\usepackage{longtable}
\usepackage{array}
\usepackage{caption}
\AtBeginDocument{%
  \providecommand\BibTeX{{%
    \normalfont B\kern-0.5em{\scshape i\kern-0.25em b}\kern-0.8em\TeX}}}

\setcopyright{acmlicensed}
\copyrightyear{2025}
\acmYear{2025}
\acmDOI{XXXXXXX.XXXXXXX}

\acmConference[Conference acronym 'XX]{Make sure to enter the correct
  conference title from your rights confirmation emai}{June 03--05,
  2018}{Woodstock, NY}
\acmISBN{978-1-4503-XXXX-X/18/06}

\input{comments} % added by JooYoung to implement comment features.

\setcopyright{none}

\begin{document}

% \title[Exploring Histogram-Based Statistical Concepts Non-Visually]{Exploring Histogram-Based Statistical Concepts Non-Visually: A Multimodal Study with Blind and Low-Vision People}
\title[Sensing the Shape of Data]{Sensing the Shape of Data: Non-Visual Exploration of Statistical Concepts in Histograms with Blind and Low-Vision Learners}

\author{Sanchita S. Kamath}
\orcid{0000-0001-6469-0360}
\affiliation{%
 \department{School of Information Sciences}
  \institution{University of Illinois Urbana-Champaign}
  \city{Champaign}
  \state{Illinois}
  \country{USA}
  \postcode{61820}
  }
\email{ssk11@illinois.edu}

\author{Omar Khan}
\orcid{0009-0005-3209-3525}
\affiliation{%
 \department{Computer Science}
  \institution{University of Illinois Urbana-Champaign}
  \city{Urbana}
  \state{Illinois}
  \country{USA}
  \postcode{61801}
  }
\email{mkhan259@illinois.edu}

\author{Aziz Zeidieh}
\orcid{0009-0000-9334-8660}
\affiliation{%
 \department{Informatics}
  \institution{University of Illinois Urbana-Champaign}
  \city{Champaign}
  \state{Illinois}
  \country{USA}
  \postcode{61820}
  }
\email{azeidi2@illinois.edu}

\author{JooYoung Seo}
\orcid{0000-0002-4064-6012}
\affiliation{%
 \department{School of Information Sciences}
  \institution{University of Illinois Urbana-Champaign}
  \city{Champaign}
  \state{Illinois}
  \country{USA}
  \postcode{61820}
  }
\email{jseo1005@illinois.edu}

\renewcommand{\shortauthors}{Kamath et al.}

\begin{abstract}
  \input{chapter/0_abstract}
\end{abstract}




\maketitle

\input{chapter/1_introduction}
\input{chapter/2_related_work}

\input{chapter/3_evaluation}
\input{chapter/4_findings}
\input{chapter/5_discussion}
\input{chapter/6_conclusion}

\begin{acks}
  \input{chapter/7_acknowledgement}
\end{acks}

\bibliographystyle{ACM-Reference-Format}
\bibliography{references/references}

\input{chapter/8_appendix}

\end{document}

%% file: comments.tex
\usepackage{todonotes}
\usepackage{soul}  % For highlighting text with \hl

%% file: chapter/0_abstract.tex
% 150 words. Must include the following components:
Statistical concepts often rely heavily on visual cues for comprehension, presenting challenges for individuals who face difficulties using visual information, such as the blind and low-vision (BLV) community. While prior work has explored making data visualizations accessible, limited research examines how BLV individuals conceptualize and learn the underlying statistical concepts these visualizations represent.
% 2. Purpose (the participants who will be studied, and the site where the research will take place)
% 3. Data collection (the type of data, the participants, and where the data will be collected)
To better understand BLV individuals' learning strategies for potentially unfamiliar statistical concepts, we conducted a within-subjects experiment with 7 BLV individuals, controlling for vision condition using blindfolds. Each participant leveraged three different non-visual representations (Swell Touch tactile graph (STGs), shaped data patterns on a refreshable display (BDPs), sonification) to understand three different statistical concepts in histograms (skewness, modality, kurtosis). We collected quantitative metrics (accuracy, completion time, self-reported confidence levels) and qualitative insights (gesture analysis) to identify participants' unique meaning-making strategies.
% 4. Results
Results revealed that the braille condition led to the most accurate results, with sonification tasks being completed the fastest. Participants demonstrated various adaptive techniques when exploring each histogram, often developing alternative mental models that helped them non-visually encode statistical visualization concepts.
% 5. Implications for target audience
Our findings reveal important implications for statistics educators and assistive technology designers, suggesting that effective learning tools must go beyond simple translation of visual information to support the unique cognitive strategies employed by BLV learners.

%% file: chapter/1_introduction.tex
\section{Introduction}
\label{sec:introduction}

% must include the following components:
% Navigation of statistical concepts via non-visual means has limited exploration 
% 
% 1. State the research problem (fact-based, objective stats on the research topic).
Statistical concepts such as skewness, modality, and kurtosis are foundational for interpreting data distributions, yet their presentation is overwhelmingly visual -- relying on charts and graphs to communicate abstract relationships and patterns~\cite{seoBornAccessibleData2024,lundgard_accessible_2022,park_impact_2022,stone_teaching_2019}. This visual dependency creates significant barriers for blind and low-vision (BLV) learners~\footnote{In this work, we refer to "learners" broadly to include both those engaged in formal education and individuals pursuing knowledge acquisition throughout their lives beyond academic environments in informal learning settings.}, who must navigate not only inaccessible representations but also the cognitive challenge of grasping these statistical ideas without visual cues~\cite{kim_exploring_2023, fan_accessibility_2023, choi_visualizing_2019}. For BLV individuals, the lack of accessible, non-visual pathways to understanding statistical concepts can hinder both confidence and data literacy~\cite{alcaraz-martinez_enhancing_2024, kay_tactile_nodate,mccallister_teaching_2001,seoBornAccessibleData2024}.
% 2. Review prior studies that have addressed the problem (some citations).                
Prior research has explored a range of non-visual representations to make statistical information accessible, including tactile graphics (STGs), Braille (BDPs), sonification, and natural language descriptions~\cite{fritz_design_1999, ali_sonify_2020, chundury_towards_2022, siu_supporting_2022, lundgard_accessible_2022, belle_alt-texify_2025}. Tactile aids can help learners physically explore the shape and spread of distributions, while sonification encodes data features such as skewness or modality into audio cues like pitch and rhythm~\cite{kay_tactile_nodate, ali_sonify_2020}. Recent advances also leverage large language models (LLMs) and interactive systems to generate accessible chart descriptions and support question answering for BLV users~\cite{seo_maidr_2024, kim_exploring_2023, alam_seechart_2023}. 
% 3. Indicate deficiencies in the studies.
Despite these innovations, most solutions focus on usability or access issues in data visualizations, often neglecting the deeper cognitive processes and learnability of statistical concepts for BLV learners~\cite{chundury_understanding_2024, wu_our_2024, sharif_understanding_2024,seoBornAccessibleData2024}. There remains a critical gap in understanding how BLV learners perceive, internalize, and build confidence in statistical concepts like skewness, modality, and kurtosis through non-visual means~\cite{godfrey_advice_2015, vita_blind_2014, mccallister_teaching_2001}.
% 4. Advance the significance of the study for particular audiences.
To address this gap, we conducted a within-subjects experiment (n=7) with BLV participants, focusing on the learnability and accessibility of skewness, modality, and kurtosis using histograms as the data visualization context. We systematically varied the representation methods -- STGs, BDPs, and sonification -- and measured participants’ confidence, completion time, and accuracy in interpreting these statistical concepts.
% 5. State the purpose statement (one purpose).
% 6. State research questions (up to three).
We chose histograms for their ability to visually convey the three aforementioned statistical concepts -- skewness, modality, and kurtosis. We randomized the order in which each concept would be presented to participants and measured participants' confidence level and understanding of each statistical concept. The following research questions (RQs) guided our investigation:
\begin{enumerate}
    \item[RQ1:] How do BLV individuals explore statistical concepts with non-visual representations?
    \item[RQ2:] How do different non-visual representations (STGs, BDPs, sonification) influence BLV individuals’ confidence, efficiency, and accuracy in learning these statistical concepts?
    \item[RQ3:] What additional support or guidance do BLV learners seek when exploring statistical concepts?
\end{enumerate}

% 7. State contributions of the study.
By foregrounding the cognitive and experiential aspects of learning skewness, modality, and kurtosis, our study contributes a nuanced understanding of how non-visual data representations can foster deeper statistical insight for BLV learners~\cite{lundgard_accessible_2022, chundury_towards_2022, siu_supporting_2022}. Our findings inform the design of accessible educational tools and pedagogical strategies, aligning with best practices in inclusive HCI and accessibility research~\cite{zhang_charta11y_2024, sharif_conveying_2023, blanco_olli_2022}.
% 8. Outline the structure of the paper.
We first situate our work within broader literature and prior investigations into pedagogical approaches to teaching statistics and efforts to make these experiences more accessible to learners with diverse abilities (Section~\ref{sec:related_work}). We then outline experimental setup and analysis techniques used for data collection (Section~\ref{sec:methods}). We move on to covering the key takeaways from our multimodal approach to teaching statistical concepts to BLV learners (Section~\ref{sec:findings}), and then discuss broader implications and of our study on statistics education for diverse learners and provide recommendations towards pedagogy that considers a multitude of approaches to teaching and learning (Section~\ref{sec:discussion}).

% From 04/08 meeting:
% Little done to understand how BLV people perceive statistical concepts - more about cognitive side and learnability. How much can BLV people understand or perceive out of it. Multimodal within subject experimentation using histogram as example - reason - histogram convery three different statistical concepts - modality, skewness and kurtosis. Randomized modality and concept and measured confidence level and understanding. Study contributed to deep understanding of how different modalities can provide perceivability of these concepts for BLV people.

%% file: chapter/2_related_work.tex
\section{Related Work}
\label{sec:related_work}

% \subsection{some sub-headings here...}

\subsection{Statistics Education: Cognitive Strategies And Learning Challenges For BLV Students}
\label{subsec:stats_ed_intro}

Understanding core statistical concepts such as skewness, modality, and kurtosis is essential for developing statistical literacy and reasoning~\cite{crd_compilation_2025, park_impact_2022, stone_teaching_2019}. For sighted learners, visualizations like histograms serve as external representations that externalize these properties, easing cognitive load and supporting pattern recognition~\cite{kang_sightation_2025, crd_compilation_2025}. In contrast, BLV learners lack visual cues, requiring alternative strategies and adaptive techniques~\cite{szafir_inclusive_2021, kim_exploring_2023, sharif_conveying_2023, mccallister_teaching_2001,seoBornAccessibleData2024, godfrey_advice_2015}. Prior research indicates that BLV individuals often rely on tactile, auditory, and linguistic cues to interpret data distributions, frequently referencing natural language queries describing visual features~\cite{kim_exploring_2023, alam_seechart_2023}.

Recent studies highlight a preference among BLV users for plain-language descriptions of statistical information, including uncertainty and distributional shapes, combined with opportunities for interactive, drill-down exploration~\cite{lundgard_accessible_2022, brown_comparing_2023, sharif_conveying_2023, smits_altgosling_2024, belle_alt-texify_2025}. Despite these advances, prior work has focused on statistical cognition that overlooks the lived experiences and unique reasoning strategies of BLV individuals~\cite{fan_accessibility_2023, oliveira_dos_santos_math_2023, chundury_understanding_2024}. This gap is critical, as BLV learners tend to develop distinct approaches to understanding and reasoning about abstract statistical properties, shaped by years of navigating a visually-oriented world~\cite{vita_blind_2014, mccallister_teaching_2001,seoBornAccessibleData2024}.

\subsection{Towards Accessible Statistics Education For The BLV Community}
\label{subsec:acccessible_stats_ed}

Efforts to make statistical visualizations accessible to BLV learners have primarily employed three non-visual modalities: tactile representations, sonification, and natural language descriptions. STGs, such as those produced with swell-form (thermoform) touch paper, allow users to physically explore the shape and spread of distributions, supporting embodied cognition and kinesthetic learning~\cite{kay_tactile_nodate, vita_blind_2014, blanco_olli_2022, stone_teaching_2019}. Swell-form paper, in particular, is widely used due to its affordability and ease of production, enabling rapid prototyping of raised-line graphics for educational settings. However, these static STGs can be limited in interactivity and may not scale well for exploring large or dynamic datasets. In addition, BDPs, which can dynamically render tactile symbols and Unicode characters, offer a flexible, time-efficient way to present data patterns~\cite{seoMAIDRMakingStatistical2024,holloway_refreshable_2024}. These devices allow users to explore different data representations programmatically, supporting more interactive engagement. However, their low resolution (typically 6 or 8 dots per cell) limits the granularity of data patterns that can be represented, posing challenges for detailed statistical analysis. Sonification encodes statistical features such as skewness or modality into auditory cues, enabling users to perceive data trends via pitch, rhythm, or volume~\cite{ali_sonify_2020, chundury_towards_2022, siu_supporting_2022, damsma_hearing_2024}. Sonification is cost-effective and adaptable, with implementations like ChartMusic~\cite{siu_supporting_2022} and VoxLens~\cite{sharif_voxlens_2022} capable of granular pattern representation. Nonetheless, sonification often requires training and may be less suitable for users with concurrent hearing impairments. Natural language descriptions, increasingly generated via AI, provide verbal summaries of statistical properties and facilitate interactive exploration~\cite{lundgard_accessible_2022, belle_alt-texify_2025}. These descriptions help bridge the gap between raw data and user understanding, especially when paired with multimodal interaction \cite{jooyoungseoTeachingVisualAccessibility2023,seo_maidr_2024,seoMAIDRMakingStatistical2024}.

\subsection{Multimodal and Embodied Approaches in Data and Statistical Literacy}
\label{subsec:multimodal_stats}

Recent innovations have moved toward integrated multimodal systems that combine tactile, auditory, and linguistic cues to support statistical understanding~\cite{zhang_charta11y_2024, seo_maidr_2024,seoMAIDRMakingStatistical2024,alcaraz-martinez_enhancing_2024, siu_supporting_2022}. For example, systems like ChartA11y~\cite{zhang_charta11y_2024}, MAIDR~\cite{seoMAIDRMakingStatistical2024,seo_maidr_2024}, and VoxLens~\cite{sharif_voxlens_2022} demonstrate how orchestrating touch, sound, and language can facilitate flexible exploration of statistical properties such as skewness and kurtosis at varying levels of granularity. These multimodal systems enable BLV users to switch seamlessly between modalities, fostering deeper understanding and confidence in interpreting data.

Despite these advances, most prior work has focused on making visualizations accessible rather than examining how these modalities support conceptual learning and reasoning about statistical concepts~\cite{wu_our_2024, chundury_understanding_2024,seoBornAccessibleData2024}. There remains a significant gap in understanding how BLV learners internalize and reason about statistical concept -- —such as skewness, modality, and kurtosis—across STGs, BDPs, and auditory (sonification), and linguistic (natural language) modalities, especially in the context of shaped data patterns (as will be covered in Section~\ref{sec:methods}).

Our work addresses these gaps by directly engaging BLV participants in the exploration of histogram-based statistical concepts, systematically comparing STGs, BDPs, and sonification. By foregrounding both accessibility and learnability, and by critically evaluating the benefits and limitations of each modality, we contribute to a growing body of research that seeks to make statistical education more inclusive, interactive, and effective for BLV learners~\cite{lundgard_accessible_2022, zhang_charta11y_2024, chundury_understanding_2024, sharif_understanding_2024, wu_our_2024,seoBornAccessibleData2024}.

%% file: chapter/3_evaluation.tex
\section{Methods}
\label{sec:methods}

To investigate how BLV individuals interpret statistical concepts represented in histograms through non-visual means, we conducted an in-person user study using STGs, BDPs, and sonification. Participants completed structured tasks assessing their understanding of skewness, modality, and kurtosis across three representations. The study design combined a within-subjects evaluation protocol (refer to Table~\ref{appendixtab:study-design}) with mixed-methods analysis, incorporating both quantitative performance metrics and qualitative behavioral observations. This section details materials, participant demographics, apparatus setup, study procedure, data collection, and analysis approach used in our study.

\subsection{Creating Materials}
\label{subsec:creating-material}
To measure non-visual interaction with histogram representations, we developed materials incorporating STG, sonification, and BDP-based data pattern representations~\footnote{Study materials can be found in the "Files" tab  \href{https://osf.io/y8eav/?view_only=8dcc168c9cff4bc585bb8055137a8c3c}{on the Open Science Framework (OSF)}.}. All materials were designed for in-person use in a portable and consistent format. STGs were produced using swell paper and enhanced with a PIAF (Pictures in a Flash) machine, yielding raised-line textures on letter-sized sheets. Sonification (non-speech auditory) stimuli were created using the MAIDR web interface~\cite{seoMAIDRMakingStatistical2024,seo_maidr_2024}, which mapped x-axis to stereo panning and y-axis value to pitch. The resulting sonification's were recorded using Audio Hijack by a member of the research team and loaded on to the Elgato Streamdeck.
BDPs were created using Unicode Braille dots to encode histogram visual patterns on a refreshable braille display, following the approach by \citet{seoMAIDRMakingStatistical2024}. We divided the data range into three segments (0–25\% low, 26–75\% medium, and 76–100\% high) and used distinct Braille cell patterns: \braille{-} for low, \braille{:} for medium, and \braille{c} for high. For example, a positively skewed pattern was encoded as: 
\braille{c}\braille{c}\braille{:}\braille{:}%
\braille{-}\braille{-}\braille{-}\braille{-}\braille{-}%
\braille{-}\braille{-}\braille{-}\braille{-}\braille{-}%
\braille{-}\braille{-}.
This Braille pattern was saved as a text file and loaded onto a 40-cell single-line refreshable Braille display for participant exploration.

To comprehensively evaluate statistical concepts in histograms, we prepared nine interventions across three non‑visual representations (resulting in 27 tasks in total) and randomized the task order (see Table~\ref{tab:intervention}). All study materials used in this study are available at \url{https://osf.io/y8eav/?view_only=8dcc168c9cff4bc585bb8055137a8c3c}.

\begin{table}[ht]
  \centering
  \caption{Nine interventions testing statistical properties, each presented through three non-visual representations, yielding 27 variations.}
  \scriptsize
  \begin{tabular}{lll}
    \toprule
    Skewness           & Modality    & Kurtosis    \\
    \midrule
    No skew            & Unimodal    & Mesokurtic  \\
    Positively skewed  & Bimodal     & Platykurtic \\
    Negatively skewed  & Multimodal  & Leptokurtic \\
    \bottomrule
  \end{tabular}
  \label{tab:intervention}
\end{table}

\subsection{Participants}
\label{subsec:participants}

With approval from the Institutional Review Board (IRB), we recruited 6 participants (P1-P6) through the mailing lists of the National Federation of the Blind (NFB)~\footnote{https://www.nfb.org/} and our final participant (P7) through convenience sampling. Participants received a \$30 Amazon gift card for participating in the one-hour, in-person study. Recruitment aimed to include individuals with varied employment and educational backgrounds. Eligibility criteria included:

\begin{itemize}
  \item Age 18 or older.
  \item Self-identify as legally blind or low-vision.
  \item Regular use of assistive technologies such as screen readers or magnifiers (e.g., JAWS, NVDA, ZoomText, VoiceOver, Magnifier).
\end{itemize}

The participant pool included four females and three males, aged 34 to 70 years ($\bar{x} = 50.00, \textit{s} = 11.96$). Six participants held graduate degrees, while one had an associate-level credential. Professional roles included engineering, teaching, self-employment, and part-time work. Table~\ref{tab:participant-table} summarizes key demographics.

\begin{table}[ht]
\centering
\scriptsize
\begin{tabular}{c c c c c c c c}
\hline
ID & Gender & Age & Education & Major & Occupation & Technology Used & Vision  \\
\hline
P1  & Female & 48  & Master's Degree & Cello Performance & Self-employed & Only Screen Reader & Blind \\
P2  & Female & 57  & Associate Degree & Liberal Arts & Part-time & Only Screen Magnifier & Low Vision \\
P3  & Female & 56  & Master's Degree & Social Work & Unemployed & Screen Reader and Screen Magnifier & Low Vision \\ 
P4  & Female & 70  & Master's Degree & Social Science & Self-employed & Screen Reader and Screen Magnifier & Low Vision \\
P5  & Male & 43  & Doctoral Degree & Chemical Engineering & Full-time & Only Screen Reader & Blind \\
P6  & Male & 42  & Associate Degree & Liberal Arts and Science & Unemployed & Screen Reader and Screen Magnifier & Blind \\
P7  & Male & 34  & Doctoral Degree & Economics & Part-time & Screen Reader and Large Font & Low Vision \\ 
\hline
\end{tabular}
\caption{Demographic characteristics of study participants.}
\label{tab:participant-table}
\end{table}

% Participants had varying experiences with vision loss, with onset ranging from birth to age 54. braille knowledge ranged from frequent use to no formal training, with several expressing interest in learning. Only one participant (P1) used a refreshable braille display. Participants primarily accessed data visualizations using screen readers, though some reported using tactile graphics or braille. None had prior experience with Swell Touch tactile or sonified histograms. Confidence in interpreting histograms varied across individuals, rated on a 7-point scale.

\subsection{Apparatus}
\label{subsec:apparatus}

The study was conducted on-site at the National Federation of the Blind (NFB) State Convention on October 17, 2024. Sessions were held in a semi-secluded area of the exhibit hall at the convention venue, with an adjacent tutorial room used for onboarding. The main setup featured a rectangular table where the participant sat on one side in a stable, armless chair. Opposite them, a researcher operated a laptop to manage stimulus presentation, log responses, and control sonification playback. A movable-arm document camera was positioned overhead to capture participants’ hand interactions during tasks. STG letter-size paper with histograms printed on them were placed in front of participants for them to be able to touch. For sonification tasks, participants wore stereo headphones delivering spatialized audio at consistent volume. Audio playback was controlled via an Elgato Streamdeck to ensure synchronized task timing and participants could touch the buttons on the hardware to replay audio clips. The BDP stimuli were rendered using a HIMS QBraille 40. All elements were arranged to provide a consistent, non-visual interaction environment across conditions.

\subsection{Procedure}
\label{subsec:procedure}

Upon arrival, participants were welcomed and briefed, followed by consent procedures. Each session began in a quiet tutorial room, where participants received verbal instruction on key statistical concepts: skewness, modality, and kurtosis. No study materials were shown at this stage. This phase allowed participants to ask clarifying questions and gain conceptual familiarity.
Participants were then escorted to the main experiment area. Before each task, they were blindfolded to eliminate visual input and standardize sensory experience. While we acknowledge the importance of preserving the experiences of BLV individuals, including the meaningful role of residual vision, we sought to eliminate variability introduced by differing levels of visual acuity. Specifically, representations such as STGs possess high visual contrast, which could be partially interpreted by participants with higher acuity. This would introduce an uneven basis for evaluating understandability, as some participants might rely on visual cues while others cannot. Since we could not reliably control for or equate this across participants, blindfolding ensured a level playing field and allowed us to isolate the effectiveness of each non-visual representations more precisely.

Seated at the interaction table, participants completed tasks across three blocks -- STGs, sonification, and BDPs -- in a counterbalanced (based on representations of delivery) within-subjects design (refer to Table~\ref{appendixtab:study-design} for details). Each block included one task per histogram property, presented in random order. At the beginning of each task, the researcher verbally introduced the property of interest (e.g., "Is the histogram positively skewed, negatively skewed, or not skewed?"). Participants were asked not to to touch the material before the task commenced with a "Ready? Go!" cue. Only for the sonification, however, participants were first (before the actual task) given a demo media file to familiarize them with the Streamdeck hardware. Participants interacted with the stimulus, then indicated completion with "Done." The researcher then recorded their verbal answer, followed by a confidence rating from 1 (not confident at all) to 7 (extremely confident). Each session concluded with a short exit interview asking participants to reflect on their experience and preferred representation type. Refer to Section~\ref{appendixtab:study-design} for the complete user study protocol.

\subsection{Data Collection}
\label{subsec:data_collection}

Participant responses and confidence ratings were logged in real time using a structured digital form. Standardized input fields minimized transcription errors across tasks. Audio and video recordings supplemented manual data entry. The document camera captured STG and BDP interactions for post hoc review, while a secondary audio device recorded verbal responses as a backup. Data were labeled with anonymized identifiers and securely stored on an encrypted research server. All media files were time-stamped and indexed to align with logged task data.

\subsection{Data Analysis}
\label{subsec:data_analysis}

We employed a mixed-methods analysis framework to examine both outcome performance and interaction behavior. 

Quantitative data included accuracy, time completion, and confidence for each task. Metrics were analyzed by representation type and histogram property to assess perceptual performance. Friedman test was used to detect overall differences among the three representations (STG, BDP and sonification) for each dependent variable: accuracy, response time, and confidence for our non-parametric within-subjects analysis. For follow-up pairwise comparisons, Wilcoxon signed-rank tests were conducted to assess differences between specific pairs of representations. Spearman’s rank correlation coefficients were calculated to examine associations between self-reported confidence and task completion time within each representation. This comprehensive analytical framework allowed us to identify not only representation-level differences, but also interaction patterns between performance metrics and user experience across modalities.
 
Qualitative analysis focused on coded interaction behaviors from the gesture analysis codebook. For each STG and BDP task, we annotated approach style, hand usage, scanning direction, and counter motions and generated codes. For each sonification task, we counted the number of repetitions and audio switches participants conducted; refer to Table~\ref{appendixtable:gesture-analysis} for a detailed overview of all codes and analyses. These codes were used to identify common strategies and how they related to successful interpretation. Open-ended comments and researcher observations were also reviewed to validate our qualitative and quantitative findings. Frequency-based analysis helped map exploration consistency across participants and conditions.

This combined approach enabled both quantitative assessment of task outcomes and qualitative insights into the processes by which BLV participants engaged with non-visual statistical graphics.

%% file: chapter/4_findings.tex
\section{Findings}
\label{sec:findings}
In this section we report patterns that emerged from participant responses and interactions during the study. The results reflect differences across representations in task performance, user confidence, and exploratory behavior. Both quantitative and qualitative analyses inform our understanding of how participants approached the interpretation of histogram properties. These results speak directly to RQ1 and RQ2, which explore how BLV individuals interpret statistical concepts using different non-visual modalities and how these modalities affect performance and experience.

\subsection{Quantitative Results - Answering RQ1 and RQ2}
\label{subsec:quant-results}
A series of non-parametric statistical tests were conducted to evaluate participant performance across the three non-visual representations. This analysis addresses RQ2 by examining how different modalities impact measurable outcomes such as accuracy, speed, and confidence.

First, since our study is partially counterbalanced, we found it imperative to test for order effect, which we found was not statistically significant ($p$ = .062, $\alpha$ = .05). This ensures the validity of within-subject comparisons across modalities in relation to RQ2. Second, we identified key metrics that included accuracy, task completion time (in seconds), and self-reported confidence (on a 7-point Likert scale). These metrics form the basis of our analysis for RQ2. A Friedman test revealed no statistically significant difference in accuracy across representations ($\chi^2$(2) = 3.429, $p$ = .180), though descriptive trends suggested higher accuracy in the BDP condition. This finding provides insight into the performance differences between representations as posed in RQ2.

Participants completed 21 tasks per representation over multiple statistical concepts. For STG, 16 tasks out of the 21 were completed accurately, yielding a success rate of 76.2\%. This high level of performance suggests that STG representations provided a relatively intuitive and interpretable means of encoding statistical information. This finding contributes to RQ1 by revealing how BLV participants engage with and make sense of STG data. Tasks involving skewness and kurtosis in particular appeared to benefit from STGs, likely due to the inherent physical affordances of STG materials. This observation speaks to RQ1 by illustrating how the characteristics of STGs align with specific statistical concepts. The sonification representation exhibited a lower accuracy rate, with participants correctly completing 14 of 21 tasks (66.7\%). While this still reflects above-chance performance, it points to potential challenges in interpreting auditory encodings of statistical patterns. This informs RQ2, as it highlights representational limitations that may impact learning. The BDPs achieved the highest accuracy, with 18 out of 21 tasks (85.7\%) completed correctly. Participants appeared to adapt quickly to BDPs, even when differentiating between subtle statistical differences like platykurtic and leptokurtic distribution. This result further develops RQ2 by demonstrating strong accuracy outcomes for BDP.

For response time, a Friedman test indicated a statistically significant difference between representations ($\chi^2$(2) = 7.000, $p$ = .030). Post-hoc Wilcoxon signed-rank tests showed that sonification tasks were completed significantly faster than STG tasks ($Z$ = 0.0, $p$ = .043), with no significant differences between STG and BDP or between BDP and sonification. These findings contribute directly to RQ2, highlighting efficiency differences across modalities.
Confidence ratings also varied significantly by representation ($\chi^2$(2) = 8.286, $p$ = .016). Participants expressed the highest confidence in STG tasks, significantly more than in sonification tasks ($Z$ = 0.0, $p$ = .018). Confidence in BDP tasks fell between the two but did not differ significantly from either. This comparative confidence analysis directly addresses RQ2 by showing how representation influences subjective certainty.
To further examine the relationship between confidence and performance, Spearman correlations were computed between confidence and response time. In the sonification condition, a significant negative correlation was found ($\rho$ = -0.712, $p$ = .047), suggesting that longer durations were associated with lower confidence. No significant correlation was observed in the STG or BDP conditions. This elaborates on RQ2 by exploring internal consistency between performance metrics.

\begin{figure}[ht]
    \centering
    \begin{minipage}[b]{0.48\textwidth}
        \centering
        \includegraphics[width=\textwidth]{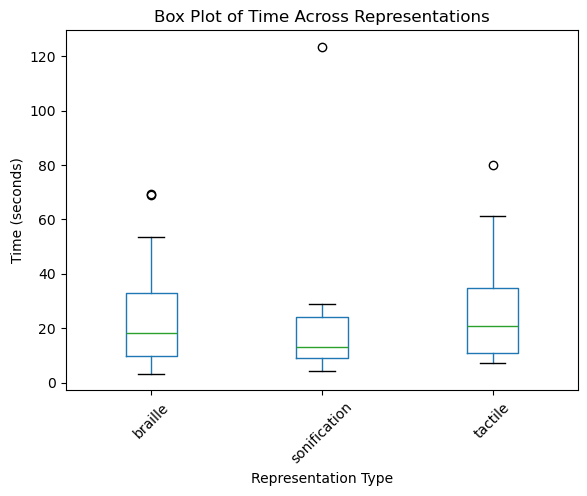}
        \caption{Task completion times across Representations}
        \label{fig:time-boxplot}
        \Description{This figure shows the distribution of task completion times for three non-visual representations: BDP, sonification, and STG. Each representation is shown as a box plot that conveys the spread and central tendency of response times in seconds. The median time for sonification tasks is the lowest, around 12 seconds, indicating that participants generally completed these tasks more quickly. BDP and STG representations both show higher medians, closer to 15 and 18 seconds respectively, and a wider range of times. Outliers are present in all three conditions, with especially long durations observed in sonification (up to 120 seconds) and STG (around 80 seconds) conditions. This suggests that while sonification enabled faster task completion on average, some participants took longer due to variability in interpretive strategies or familiarity.}
    \end{minipage}
    \hfill
    \begin{minipage}[b]{0.465\textwidth}
        \centering
        \includegraphics[width=\textwidth]{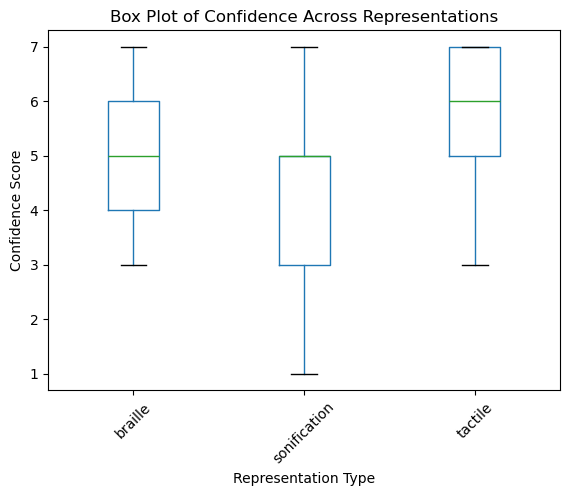}
        \caption{Confidence ratings across Representations}
        \label{fig:confidence-boxplot}
        \Description{This figure displays participants’ self-reported confidence scores, rated on a 7-point Likert scale, across the three representation types. The STG condition yielded the highest median confidence score, around 6, with responses tightly clustered near the upper end of the scale, suggesting strong and consistent user confidence. The BDP condition also shows relatively high confidence, with a median of 5 and values ranging from 3 to 7. In contrast, the sonification condition exhibits both a lower median (also around 5) and the widest range of responses—from as low as 1 to as high as 7—indicating greater variability and uncertainty among participants.}
    \end{minipage}
\end{figure}

Figures~\ref{fig:time-boxplot} and~\ref{fig:confidence-boxplot} show that sonification tasks generally resulted in faster responses and more confident responses for STG tasks. The wider spread and outliers in sonification and BDP conditions highlight the increased variability and initial challenge of interpreting data through less familiar representations.

\subsection{Qualitative Observations - Answering RQ1}
\label{subsec:qual-results}
This result primarily addresses RQ1 by detailing how BLV participants physically and cognitively explored statistical representations through touch and sound. 
Participants engaged actively with both STG and sonification representations, using a range of gestures to interpret statistical information. STG exploration typically involved contour tracing to discern shape, trend direction, and comparative features, such as identifying peaks or skew. In sonification tasks, participants engaged through active key pressing, listening carefully to pitch shifts and rhythm patterns to infer modality and skewness. Gestural behavior often mirrored the structure of statistical concepts. Skewness was embodied through directional leaning or one-sided tracing; kurtosis was interpreted through sharp or broad sweeps that mimicked distribution ``peakedness''; and representation was reflected in single or multiple passes across raised peaks. These movements reflected participants' internal conceptual models of distribution shape, rendered physically through touch or navigated audibly through key-based audio scans. Two dominant exploration strategies emerged: whole-to-part (w2p) and part-to-whole (p2w). Participants used these flexibly, often depending on their familiarity with the task. Left-to-right scanning was most common, particularly in interpreting modality, but was frequently supplemented by top-to-bottom or center-outward patterns when assessing kurtosis. Directional sequences like \texttt{l[2], r[1], u[1]} showed participants’ methodical approach to locating and verifying features across axes. Find all our developed codes in Table \ref{appendixtable:gesture-acronyms} and the complete gesture analysis in Table \ref{appendixtable:gesture-analysis}.

While the dominant hand (typically right) was used primarily for focused exploration, several tasks involved bi-manual coordination, especially when symmetry or comparative assessment was required. Simultaneous use of both hands was observed in identifying bimodal distributions or when distinguishing central vs. peripheral features. This behavior highlighted embodied spatial reasoning and tactile memory strategies. Participants often engaged in repeated left-right or circular motions prior to decision-making. These counter-movements functioned as hypothesis-testing behavior, allowing participants to confirm or refine their interpretations before answering. Tasks involving uncertainty or complex distributions saw a greater number of iterative gestures, suggesting that exploration cycles served as an embodied feedback loop. Post-task reflections and researcher notes provided additional insight into participant experience. While minor administrative interruptions were noted by a few participants, most demonstrated persistence and strategic engagement throughout. These comments supported observed correlations between gesture confidence, clarity, and task success.

These findings demonstrate how BLV participants utilize multisensory strategies, gestural routines, and embodied inference to access and interpret complex statistical information. These strategies reflect not only physical adaptation but conceptual reasoning embedded in action.

\subsection{Participant Reflections' Analysis - Answering RQ3}
\label{subsec:exit-interview}
This finding addresses RQ3, focusing on the types of support and guidance that participants expressed a need for, during and after their engagement with non-visual statistical materials. The insights include preferences for clearer explanations, repeated terms, and procedural cues that enhance confidence and comprehension.
Participants occasionally verbalized their interpretive processes, reflecting growing understanding or ongoing uncertainty. Exit interviews and in-task reflections highlighted a desire for contextual learning alongside sensory materials. Several participants requested more verbal context and clarification about statistical terms like ``mesokurtic'' or ``bimodal'' before being asked to apply them. Others expressed a need for step-by-step orientation cues, particularly when working with less familiar representations such as sonification or BDPs. Finally, most participants requested repetition of terms  - finding it difficult to correlate statistical terms to concepts itself, which prompted the researchers to ask participants to explain the concepts in their own words rather than using jargon. Participants also emphasized the need to know what to look for — e.g., how many peaks to expect, or how sharpness correlates with kurtosis — suggesting that procedural clarity enhances confidence. These findings indicate that access to non-visual representations must be accompanied by descriptions in order to support independent reasoning and reduce uncertainty in interpretation.

% objective findings corresponding to each research question defined in Introduction section.

%% file: chapter/5_discussion.tex
\section{Discussion}
\label{sec:discussion}

\subsection{BLV Learners' Exploration of Statistical Concepts}
\label{subsec:blv_exploration}

Our findings demonstrate that BLV learners can meaningfully engage with core statistical concepts -- skewness, modality, and kurtosis -- when provided with thoughtfully designed non-visual representations. Participants employed a range of adaptive strategies, such as directional scanning, repetitive touch, and sequential audio playback, to interpret distributional patterns. Rather than attempting to reconstruct a visual image, participants focused on localized perception, identifying salient features like peaks, outliers, and spread through tactile and auditory cues. This echoes prior work showing that BLV learners develop unique, embodied approaches to statistical reasoning~\cite{kim_exploring_2023, sharif_conveying_2023, vita_blind_2014, mccallister_teaching_2001}. Furthermore, our results reinforce the idea that statistical interpretation is not inherently visual, but can be achieved through structured multi-sensory interaction, supporting calls for a broader, more inclusive understanding of statistical cognition~\cite{lundgard_accessible_2022, chundury_towards_2022,seoBornAccessibleData2024,jooyoungseoTeachingVisualAccessibility2023}.

\subsection{Multimodal Access to Statistics Education}  
\label{subsec:multimodal_education}

Each non-visual representation offered distinct affordances and challenges for learning statistical concepts. STGs were associated with higher confidence and familiarity, likely due to their immediate spatial layout and alignment with established pedagogical practices~\cite{kay_tactile_nodate, stone_teaching_2019, blanco_olli_2022}. Sonification, while initially less intuitive, enabled rapid exploration and, with practice, supported nuanced understanding of features like skewness and modality~\cite{ali_sonify_2020, siu_supporting_2022}. BDP yielded the highest accuracy for some participants, but comfort and interpretability varied, highlighting the importance of individual preference and prior experience. These findings align with recent literature advocating for flexible approaches to accessible data visualization~\cite{zhang_charta11y_2024, seo_maidr_2024,alcaraz-martinez_enhancing_2024,seoMAIDRMakingStatistical2024}. Critically, our results suggest that no single modality is universally optimal; rather, providing multiple access points enables BLV learners to select and combine strategies that best support their own sensemaking~\cite{lundgard_accessible_2022, chundury_towards_2022,seoMAIDRMakingStatistical2024,seo_maidr_2024,jooyoungseoTeachingVisualAccessibility2023}.

\subsection{Augmenting Statistics' Learnability with and for BLV Learners}
\label{subsec:augmenting_learning}

While non-visual representations can render statistical data accessible, our study highlights that accessibility alone does not guarantee learnability. Participants expressed a need for additional guidance, such as clearer explanations, procedural cues, and interpretive strategies to support their reasoning about abstract statistical properties. This finding echoes calls in the literature for instructional design that goes beyond access, foregrounding orientation and sensemaking~\cite{sharif_conveying_2023, lundgard_accessible_2022, morash_guiding_2015}. Interactive tutorials, layered feedback, and verbal prompts could scaffold the process of interpreting features like peak width or asymmetry, helping BLV learners build both confidence and accuracy. Our results suggest that future accessible statistics education should integrate non-visual representations with explicit pedagogical supports, co-designed with BLV learners, to foster deeper conceptual understanding~\cite{wu_our_2024, kim_exploring_2023, mccallister_teaching_2001}.

% deep interpretation of the findings, and must include the following components:
% 1) a summary of the findings,
% 2) a discussion of the findings in relation to the research questions,
% 3) a discussion of the findings in relation to the literature.

\subsection{Limitations and Future Directions}
\label{subsec:limitations_and_future_directions}

% limitations of the study, and implications for future research.
% \begin{itemize}
%     \item Loud, crowded venue; did our best to minimize disruptions, but still could've played a factor in participant interactions with interventions
%     \item Limited to histograms, may not carry over for all continuous data visualizations 
% \end{itemize}

While this study offers insight into how BLV individuals engage with non-visual representations of statistical distributions, several limitations must be acknowledged. First, the study was conducted in-person at a national convention, where ambient noise and occasional interruptions were present despite efforts to create a controlled and semi-secluded testing area. Although participants appeared focused and completed all tasks, the loud and crowded environment may have introduced variability in concentration or affected the clarity of sonified cues and verbal instructions. Second, the study focused exclusively on histograms as a representative continuous data visualization. While histograms encompass multiple statistical properties (e.g., shape, spread, modality), the findings may not generalize to other types of continuous data representations such as density plots, box plots, or line graphs. Future work should investigate whether the strategies and affordances observed here translate to other formats that encode similar statistical information but through different visual and spatial conventions.
Third, while the participant pool reflected a range of ages, education levels, and vision types, the sample size was relatively small and drawn primarily from individuals already engaged with assistive technology communities. As such, findings may reflect a subset of BLV users who are more familiar with tactile or auditory interfaces than the general population. This limits the generalizability of results to novice users or those with different educational or technological backgrounds. Broader sampling across settings and levels of statistical familiarity would be valuable in assessing the accessibility of non-visual representations for more diverse user groups.
Finally, while the study design incorporated counterbalancing across both representations and statistical properties, not all possible combinations were represented across participants. However, the absence of a few representation-property pairings may limit the interpretability of specific interaction effects. Future iterations of the study will expand the design to ensure full factorial coverage and strengthen comparative analyses. Future research should expand to a broader range of statistical visualizations and recruit a more diverse sample of BLV learners, including those with less experience in tactile or auditory interfaces. Additionally, future studies should explore the integration of multimodal representations with interactive, adaptive instructional supports, leveraging recent advances in AI-driven chart description and question answering~\cite{seo_maidr_2024, belle_alt-texify_2025, alam_seechart_2023,seoMAIDRMakingStatistical2024}. By centering both accessibility and learnability, and by co-designing with BLV learners, the field can move toward more inclusive, effective, and empowering statistics education for all~\cite{zhang_charta11y_2024, sharif_understanding_2024, wu_our_2024,seoBornAccessibleData2024,jooyoungseoTeachingVisualAccessibility2023}.

%% file: chapter/6_conclusion.tex
\section{Conclusion}
\label{sec:conclusion}

This work highlights a crucial shift in how we conceptualize data accessibility and the learning of statistical concepts, not merely as a technical challenge of converting visuals into tactile or auditory formats, but as a fundamentally cognitive process grounded in how people learn, reason, and build meaning through their senses.
Among the three formats, we observed that the BDPs yielded the highest accuracy and that self-reported confidence ratings were favorable for STGs and task completion times were least for sonification. In our qualitative analysis of gesture interactions with the STGs, we found that participants used a mix of whole-to-part (w2p) and part-to-whole (p2w) exploration techniques. 
Our findings reveal that BLV learners do not simply "adapt" to the absence of visuals; they engage statistical concepts through embodied, exploratory strategies that reflect their own mental models and lived experience.

Therefore, we acknowledge that some statistical concepts are not just visually self-explanatory but "visually privileged"; their translation demands more than sensory substitution, which is what we hope to focus on further, and encourage other producers of accessible data to do. 
We hope our contribution serves as the catalyst for more work to explore and advance BLV learners' perception of statistical concepts, especially as it pertains to enhancing the perceptibility and accuracy in identification of statistical concepts depicted visually. In reframing access to statistical reasoning as an experiential and interpretive process, this work opens new directions: for designing tools that prioritize user cognition over data fidelity, for curricula that center BLV learners as producers and co-generators of knowledge, and for research that embraces complexity over accommodation. In short, this study is not just about making histograms accessible, it's about rethinking what it means to teach, to learn, and to know, in the absence of visual input.

%% file: chapter/7_acknowledgement.tex
The authors wish to express their gratitude for the support and contributions that have made this research possible. Detailed acknowledgments and disclosures of funding sources and collaborations will be provided upon completion of the double-blind review process to maintain the integrity of the review.

%% file: chapter/8_appendix.tex
\appendix

\section{Counterbalancing}
\label{appendixsec:counterbalancing}
Each participant was exposed to all three representation modalities — STG, sonification, and BDP — in different orders. Likewise, the statistical properties under investigation — skewness, kurtosis, and modality — were presented in varied sequences across participants. To maintain experimental control while preserving feasibility within the study’s scope, we adopted a partial counterbalancing approach. While each participant encountered all three representations and all three statistical properties, not all possible pairwise combinations were presented across the sample. This selective balancing was intentionally applied to prioritize depth of engagement within conditions while still distributing order effects across the cohort. The approach offers a practical compromise between counterbalancing full factorial complexity and experimental compromise, setting the stage for targeted exploration in future expanded protocols.

\begin{center}
\scriptsize
\setlength{\tabcolsep}{4pt}
\renewcommand{\arraystretch}{1.1}

\begin{longtable}{ll>{\raggedright\arraybackslash}p{3.2cm}l>{\raggedright\arraybackslash}p{3.2cm}>{\raggedright\arraybackslash}p{2cm}l}
\caption{Study design matrix}
\label{appendixtab:study-design}\\
\toprule
Participant & Essential\_or\_Additional & Representation\_Order & Representation & Stat\_Property\_Order & Stat\_Property & Level \\
\midrule
\endfirsthead
\caption[]{Study design matrix (continued)} \\
\toprule
Participant & Essential\_or\_Additional & Representation\_Order & Representation & Stat\_Property\_Order & Stat\_Property & Level \\
\midrule
\endhead
\midrule
\multicolumn{7}{r}{{Continued on next page}} \\
\midrule
\endfoot
\bottomrule
\endlastfoot
P1 & Essential & STG - sonification - BDP &  STG & skewness - kurtosis - modality & skewness & no \\
P1 &  Essential & STG - sonification - BDP & STG & skewness - kurtosis - modality & kurtosis & meso \\
 P1 &Essential & STG - sonification - BDP &STG & skewness - kurtosis - modality &      modality &      uni \\
 P1 &Essential & STG - sonification - BDP &   sonification & skewness - kurtosis - modality &      skewness & positive \\
 P1 &Essential & STG - sonification - BDP &   sonification & skewness - kurtosis - modality &      kurtosis &    lepto \\
 P1 &Essential & STG - sonification - BDP &   sonification & skewness - kurtosis - modality &      modality &bi \\
 P1 &Essential & STG - sonification - BDP &BDP & skewness - kurtosis - modality &      skewness & negative \\
 P1 &Essential & STG - sonification - BDP &BDP & skewness - kurtosis - modality &      kurtosis &    platy \\
 P1 &Essential & STG - sonification - BDP &BDP & skewness - kurtosis - modality &      modality &    multi \\
 P2 &Essential & STG - BDP - sonification &STG & skewness - modality - kurtosis &      skewness & positive \\
 P2 &Essential & STG - BDP - sonification &STG & skewness - modality - kurtosis &      modality &bi \\
 P2 &Essential & STG - BDP - sonification &STG & skewness - modality - kurtosis &      kurtosis &    lepto \\
 P2 &Essential & STG - BDP - sonification &BDP & skewness - modality - kurtosis &      skewness & negative \\
 P2 &Essential & STG - BDP - sonification &BDP & skewness - modality - kurtosis &      modality &    multi \\
 P2 &Essential & STG - BDP - sonification &BDP & skewness - modality - kurtosis &      kurtosis &    platy \\
 P2 &Essential & STG - BDP - sonification &   sonification & skewness - modality - kurtosis &      skewness &no \\
 P2 &Essential & STG - BDP - sonification &   sonification & skewness - modality - kurtosis &      modality &      uni \\
 P2 &Essential & STG - BDP - sonification &   sonification & skewness - modality - kurtosis &      kurtosis &     meso \\
 P3 &Essential & sonification - STG - BDP &   sonification & kurtosis - skewness - modality &      kurtosis &    platy \\
 P3 &Essential & sonification - STG - BDP &   sonification & kurtosis - skewness - modality &      skewness & negative \\
 P3 &Essential & sonification - STG - BDP &   sonification & kurtosis - skewness - modality &      modality &    multi \\
 P3 &Essential & sonification - STG - BDP &STG & kurtosis - skewness - modality &      kurtosis &     meso \\
 P3 &Essential & sonification - STG - BDP &STG & kurtosis - skewness - modality &      skewness &no \\
 P3 &Essential & sonification - STG - BDP &STG & kurtosis - skewness - modality &      modality &      uni \\
 P3 &Essential & sonification - STG - BDP &BDP & kurtosis - skewness - modality &      kurtosis &    lepto \\
 P3 &Essential & sonification - STG - BDP &BDP & kurtosis - skewness - modality &      skewness & positive \\
 P3 &Essential & sonification - STG - BDP &BDP & kurtosis - skewness - modality &      modality &bi \\
 P4 &Essential & sonification - BDP - STG &   sonification & kurtosis - modality - skewness &      kurtosis &     meso \\
 P4 &Essential & sonification - BDP - STG &   sonification & kurtosis - modality - skewness &      modality &      uni \\
 P4 &Essential & sonification - BDP - STG &   sonification & kurtosis - modality - skewness &      skewness &no \\
 P4 &Essential & sonification - BDP - STG &BDP & kurtosis - modality - skewness &      kurtosis &    platy \\
 P4 &Essential & sonification - BDP - STG &BDP & kurtosis - modality - skewness &      modality &    multi \\
 P4 &Essential & sonification - BDP - STG &BDP & kurtosis - modality - skewness &      skewness & negative \\
 P4 &Essential & sonification - BDP - STG &STG & kurtosis - modality - skewness &      kurtosis &    lepto \\
 P4 &Essential & sonification - BDP - STG &STG & kurtosis - modality - skewness &      modality &bi \\
 P4 &Essential & sonification - BDP - STG &STG & kurtosis - modality - skewness &      skewness & positive \\
 P5 &Essential & BDP - STG - sonification &BDP & modality - skewness - kurtosis &      modality &bi \\
 P5 &Essential & BDP - STG - sonification &BDP & modality - skewness - kurtosis &      skewness & positive \\
 P5 &Essential & BDP - STG - sonification &BDP & modality - skewness - kurtosis &      kurtosis &    lepto \\
 P5 &Essential & BDP - STG - sonification &STG & modality - skewness - kurtosis &      modality &      uni \\
 P5 &Essential & BDP - STG - sonification &STG & modality - skewness - kurtosis &      skewness &no \\
 P5 &Essential & BDP - STG - sonification &STG & modality - skewness - kurtosis &      kurtosis &     meso \\
 P5 &Essential & BDP - STG - sonification &   sonification & modality - skewness - kurtosis &      modality &    multi \\
 P5 &Essential & BDP - STG - sonification &   sonification & modality - skewness - kurtosis &      skewness & negative \\
 P5 &Essential & BDP - STG - sonification &   sonification & modality - skewness - kurtosis &      kurtosis &    platy \\
 P6 &Essential & BDP - sonification - STG &BDP & modality - kurtosis - skewness &      modality &    multi \\
 P6 &Essential & BDP - sonification - STG &BDP & modality - kurtosis - skewness &      kurtosis &    platy \\
 P6 &Essential & BDP - sonification - STG &BDP & modality - kurtosis - skewness &      skewness & negative \\
 P6 &Essential & BDP - sonification - STG &   sonification & modality - kurtosis - skewness &      modality &bi \\
 P6 &Essential & BDP - sonification - STG &   sonification & modality - kurtosis - skewness &      kurtosis &    lepto \\
 P6 &Essential & BDP - sonification - STG &   sonification & modality - kurtosis - skewness &      skewness & positive \\
 P6 &Essential & BDP - sonification - STG &STG & modality - kurtosis - skewness &      modality &      uni \\
 P6 &Essential & BDP - sonification - STG &STG & modality - kurtosis - skewness &      kurtosis &     meso \\
 P6 &Essential & BDP - sonification - STG &STG & modality - kurtosis - skewness &      skewness &no \\
 P7 & Essential & STG - sonification - BDP &  STG & skewness - kurtosis - modality & skewness & no \\
P7 &  Essential & STG - sonification - BDP & STG & skewness - kurtosis - modality & kurtosis & meso \\
 P7 &Essential & STG - sonification - BDP &STG & skewness - kurtosis - modality &      modality &      uni \\
 P7 &Essential & STG - sonification - BDP &   sonification & skewness - kurtosis - modality &      skewness & positive \\
 P7 &Essential & STG - sonification - BDP &   sonification & skewness - kurtosis - modality &      kurtosis &    lepto \\
 P7 &Essential & STG - sonification - BDP &   sonification & skewness - kurtosis - modality &      modality &bi \\
 P7 &Essential & STG - sonification - BDP &BDP & skewness - kurtosis - modality &      skewness & negative \\
 P7 &Essential & STG - sonification - BDP &BDP & skewness - kurtosis - modality &      kurtosis &    platy \\
 P7 &Essential & STG - sonification - BDP &BDP & skewness - kurtosis - modality &      modality &    multi \\
\end{longtable}
\end{center}

\section{Participant Data}
\label{appendixsec:participant-data}

\begin{center}
\scriptsize
\begin{longtable}{llllrr}
\caption{Participant Data about Accuracy and Confidence} \\
\toprule
ParticipantID & Representation & Stat \_Property & Level & Accuracy & Confidence \\
\midrule
\endfirsthead

\toprule
Participant & Representation & Stat \_Property & Level & accuracy & confidence \\
\midrule
\endhead

\bottomrule
\endfoot

P1 & STG & skewness & no & 1 & 7 \\
P1 & STG & kurtosis & meso & 1 & 7 \\
P1 & STG & modality & uni & 0 & 7 \\
P1 & sonification & skewness & positive & 1 & 3 \\
P1 & sonification & kurtosis & lepto & 1 & 3 \\
P1 & sonification & modality & bi & 0 & 5 \\
P1 & BDP & skewness & negative & 1 & 4 \\
P1 & BDP & kurtosis & platy & 1 & 6 \\
P1 & BDP & modality & multi & 1 & 6 \\
P2 & STG & skewness & positive & 1 & 6 \\
P2 & STG & modality & bi & 1 & 7 \\
P2 & STG & kurtosis & lepto & 1 & 6 \\
P2 & BDP & skewness & negative & 1 & 4 \\
P2 & BDP & modality & multi & 1 & 6 \\
P2 & BDP & kurtosis & platy & 1 & 3 \\
P2 & sonification & skewness & no & 0 & 5 \\
P2 & sonification & modality & uni & 1 & 5 \\
P2 & sonification & kurtosis & meso & 1 & 5 \\
P3 & sonification & kurtosis & platy & 0 & 5 \\
P3 & sonification & skewness & negative & 1 & 4 \\
P3 & sonification & modality & multi & 1 & 7 \\
P3 & STG & kurtosis & meso & 1 & 4 \\
P3 & STG & skewness & no & 1 & 7 \\
P3 & STG & modality & uni & 0 & 5 \\
P3 & BDP & kurtosis & lepto & 0 & 7 \\
P3 & BDP & skewness & positive & 1 & 4 \\
P3 & BDP & modality & bi & 1 & 7 \\
P4 & sonification & kurtosis & meso & 1 & 1 \\
P4 & sonification & modality & uni & 0 & 2 \\
P4 & sonification & skewness & no & 1 & 6 \\
P4 & BDP & kurtosis & platy & 1 & 4 \\
P4 & BDP & modality & multi & 0 & 3 \\
P4 & BDP & skewness & negative & 1 & 7 \\
P4 & STG & kurtosis & lepto & 1 & 6 \\
P4 & STG & modality & bi & 1 & 7 \\
P4 & STG & skewness & positive & 0 & 7 \\
P5 & BDP & modality & bi & 1 & 3 \\
P5 & BDP & skewness & positive & 1 & 3 \\
P5 & BDP & kurtosis & lepto & 1 & 5 \\
P5 & STG & modality & uni & 0 & 4 \\
P5 & STG & skewness & no & 0 & 3 \\
P5 & STG & kurtosis & meso & 1 & 3 \\
P5 & sonification & modality & multi & 0 & 3 \\
P5 & sonification & skewness & negative & 1 & 5 \\
P5 & sonification & kurtosis & platy & 0 & 5 \\
P6 & BDP & modality & multi & 1 & 3 \\
P6 & BDP & kurtosis & platy & 1 & 7 \\
P6 & BDP & skewness & negative & 1 & 6 \\
P6 & sonification & modality & bi & 1 & 5 \\
P6 & sonification & kurtosis & lepto & 0 & 4 \\
P6 & sonification & skewness & positive & 1 & 6 \\
P6 & STG & modality & uni & 1 & 7 \\
P6 & STG & kurtosis & meso & 1 & 5 \\
P6 & STG & skewness & no & 1 & 6 \\
P7 & STG & skewness & no & 1 & 5 \\
P7 & STG & kurtosis & meso & 1 & 6 \\
P7 & STG & modality & uni & 1 & 4 \\
P7 & sonification & skewness & positive & 1 & 3 \\
P7 & sonification & kurtosis & lepto & 1 & 6 \\
P7 & sonification & modality & bi & 1 & 7 \\
P7 & BDP & skewness & negative & 1 & 6 \\
P7 & BDP & kurtosis & platy & 0 & 4 \\
P7 & BDP & modality & multi & 1 & 7 \\
\end{longtable}
\label{appendixtable:participant-data}
\end{center}

\section{Qualitative Analysis}
\label{appendixsec:appendix-qualanalysis}

\subsection{Codes}
\label{appendixsubsec:codes}

\begin{table}[H]
\centering
\scriptsize
\caption{List of gesture and direction acronyms used in analysis.}
\label{appendixtable:gesture-acronyms}
\begin{tabular}{lll}
\toprule
Acronym & Definition\\
\midrule
p2w & part to whole \\
w2p & whole to part \\
l2r & left to right \\
r2l & right to left \\
u2d & up to down \\
d2u & down to up \\
pu2d & chose a part and then went up and down on it \\
lrci & scan from ends toward center \\
lrco & scan from center toward ends \\
crl & center to right then left \\
clr & center to left then right \\
l[x] & left-to-right in sonification, x = repetitions \\
r[x] & right-to-left in sonification, x = repetitions \\
u[x] & upward movement x times \\
d[x] & downward movement x times \\
s1 & location on path \\
s2 & motion in space \\
s3 & source of path \\
s4 & end of path \\
s5 & re-traversing same area on path \\
s6 & moving forward on path \\
\bottomrule
\end{tabular}
\end{table}

\subsection{Gesture Analysis}
\label{appendixsubsec:gesture-analysis}

\scriptsize
\begin{longtable}{p{0.5cm}p{0.7cm}p{0.9cm}p{0.9cm}p{0.8cm}p{0.6cm}p{2.5cm}p{0.5cm}p{0.5cm}p{3cm}}
\caption{Gesture Analysis Codebook} \\
\hline
\textbf{PID} & \textbf{Rep.} & \textbf{Stat Type} & \textbf{Stat Prop.} & \textbf{Approach} & \textbf{Hands} & \textbf{Direction Sequence} & \textbf{Count} & \textbf{Motion} & \textbf{Comments} \\
\hline
\endfirsthead

\multicolumn{10}{c}%
{{\bfseries \tablename\ \thetable{} -- continued from previous page}} \\
\hline
\textbf{PID} & \textbf{Rep.} & \textbf{Stat Type} & \textbf{Stat Prop.} & \textbf{Approach} & \textbf{Hands} & \textbf{Direction Sequence} & \textbf{Count} & \textbf{Motion} & \textbf{Comments} \\
\hline
\endhead

\hline \multicolumn{10}{|r|}{{Continued on next page}} \\ \hline
\endfoot

\hline
\endlastfoot

P1 & STG & skewness & no & p2w & R & l[1],r[1] & 2 & s1,s5 & went from l2r, r2l repeatedly before giving the answer \\
\hline
P1 & STG & kurtosis & meso & p2w & R & l[1], r[1], u[7], r[1],d[3] & 13 & s1,s5 & kept feeling the graph swiftly and feeling the white space to check for 'pointy' \\
\hline
P1 & STG & modality & uni & p2w & R & l[2], u[1] & 3 & s1,s5 & went only l2r, got confused between separate multiple peaks and singular frequency peaks from each bar, hence answered wrong. \\
\hline
P1 & sonification & skewness & positive & l2r & -- & l[2],r[1],l[1] & 4 & -- & there were some admin snags, but prefered l2r \\
\hline
P1 & sonification & kurtosis & lepto & l2r & -- & l[1], l[3] & 4 & -- & played once before the question was asked, and then three times after the question was asked, one after the other, without letting the clip play fully \\
\hline
P1 & sonification & modality & bi & l2r & -- & l[2] & 2 & -- & was very confident, did not play the r2l at all \\
\hline
P1 & BDP & skewness & negative & p2w & R & l[4] & 4 & s1,s5 & did not stay on the BDP display line, considered the cursor routing part of the display as part of graph \\
\hline
P1 & BDP & kurtosis & platy & l2r & R & l[3] & 3 & s1,s5 & still went over to the cursor routing area \\
\hline
P1 & BDP & modality & multi & none & R & r[1], l[1] & 2 & s1,s5 & started from the center this time, then went left and then from left to the center, back to the beginning point, before following normal pattern. did not go the cursor routing area this time \\
\hline
P2 & STG & skewness & positive & w2p & L/R & u[2], l[1], u[1] & 4 & s2, s1,s5 & touched the border (using the border to orient themselves) with their right hand and graph with the left then used both hand to feel the graph \\
\hline
P2 & STG & modality & bi & w2p & L/R & d[1], r[1] & 2 & s2, s5, s1 & used both hands beginning from the center with more reliance on left to find the peaks \\
\hline
P2 & STG & kurtosis & lepto & p2w & L/R & u[1] & 1 & s2,s5 & looked around the plot with left hand and found the pointy bar \\
\hline
P2 & BDP & skewness & negative & w2p & L/R & lrco[1], l[1] & 2 & s1,s5 & felt around the graph from the center to the sides, found the BDP line almost immediately \\
\hline
P2 & BDP & modality & multi & p2w & L/R & l[2] & 2 & s1,s5 & both hands were feeling the area at the same time, with left closely following the right \\
\hline
P2 & BDP & kurtosis & platy & p2w & L/R & l[1], r[1], l[1] & 3 & s1,s5 & same strategy as the one above \\
\hline
P2 & sonification & skewness & no & l2r & -- & l[2],r[2],l[2],r[1] & 7 & -- & did not let the whole thing play except the last right \\
\hline
P2 & sonification & modality & uni & none & -- & l[2],r[2] & 4 & -- & played the whole clip \\
\hline
P2 & sonification & kurtosis & meso & none & -- & l[2], r[2] & 4 & -- & let the whole clip play, but waited for a slight moment before playing the fifth clip, but said done instead \\
\hline
P3 & sonification & kurtosis & platy & none & -- & l[1], r[1] & 2 & -- & reasoned their argument, but did not listen to the clip for a very long time \\
\hline
P3 & sonification & skewness & negative & none & -- & l[1],r[1] & 2 & -- & mixed up their answer, but caught the concept. did not wait for the second clip to play completely before answering \\
\hline
P3 & sonification & modality & multi & l2r & -- & l[1] & 1 & -- & played only one clip and did not wait for the whole clip to finish before they said done \\
\hline
P3 & STG & kurtosis & meso & p2w & L & l[1] & 1 & s1 & did not even do the whole graph, only felt a portion of it \\
\hline
P3 & STG & skewness & no & p2w & L & l[1] & 1 & s1 & saw the whole graph, but did not take a lot of time to ponder, forgot to answer after done \\
\hline
P3 & STG & modality & uni & p2w & L & l[1],r[1] & 2 & s1,s5 & for all three STGs, did a tippy-tippy touch with single finger instead of using all fingers and palm. Futher, they also got confused between singular frequency peaks and actual peaks \\
\hline
P3 & BDP & kurtosis & lepto & p2w & L & l[1],r[1] & 2 & s1,s5 & did the brush touch using single finger for all BDP, but in this instance actually brushed over and did not feel the high point at all \\
\hline
P3 & BDP & skewness & positive & p2w & L & l[1] & 1 & s1 & pondered a lot when asked a follow-up question and gave the right answer/ confused with the terminology \\
\hline
P3 & BDP & modality & bi & p2w & L & l[1] & 1 & s1 & was very confident in answering, felt the peaks well this time \\
\hline
P4 & sonification & kurtosis & meso & none & -- & l[1],r[1],l[1],r[1] & 4 & -- & let whole clips play and was very speculative \\
\hline
P4 & sonification & modality & uni & none & -- & l[1], r[1], l[1], r[1], l[1], r[1] & 6 & -- & let whole clips play and seemed to second guess their answers \\
\hline
P4 & sonification & skewness & no & none & -- & r[1], l[1], l[1], l[1], r[1], r[1], r[1], l[1] & 8 & -- & \\
\hline
P4 & BDP & kurtosis & platy & p2w & L/R & l[1], r[1], lrco[1], lrci[1], lrco[1], lrci[1], lrco[1], lrci[1], lrco[1], lrci[1], lrco[1], lrci[1], lrco[1] & 13 & s1,s5 & touched over and over again center to outside/inside \\
\hline
P4 & BDP & modality & multi & p2w & L/R & crl[1], lrco[1], lrci[1], l[1], lrco[1], l[1], lrco[1], lrci[1], lrco[1], lrci[1], lrco[1], lrci[1], lrco[1], lrci[1], lrco[1], lrci[1] & 16 & s1,s5 & started from the center with left hand, moved towards left and then started using right hand which followed \\
\hline
P4 & BDP & skewness & negative & p2w & L & crl[1],l[1],r[1] & 3 & s1,s5 & seemed much more confident in this round \\
\hline
P4 & STG & kurtosis & lepto & w2p & L/R & pu2d[3] & 3 & s1,s2 & with left to right strokes when going pu2d \\
\hline
P4 & STG & modality & bi & p2w & L/R & u[1],r[2],u[2] d[3] & 8 & s1,s5 & after coming to the center of the trough between the two peaks, did not move on to the second peak at all. instead kept feeling the first peak. They did not touch the second peak even once to ignore it. \\
\hline
P4 & STG & skewness & positive & p2w & R & l[3] & 3 & s2,s1,s5 & seemed very confident and did not second guess as much. The first thing they touched was the blank space \\
\hline
P5 & BDP & modality & bi & p2w & L & l[1] & 1 & s1 & \\
\hline
P5 & BDP & skewness & positive & p2w & L & crl[2],l[1] & 3 & s1,s5 & somehow well touched every point and still confidently got it wrong despite prompting from the PI \\
\hline
P5 & BDP & kurtosis & lepto & p2w & L/R & l[1] & 1 & s1 & used the left hand till the center and then used the right hand for the rest of the way \\
\hline
P5 & STG & modality & uni & p2w & L/R & l[1],r[1] & 2 & s1,s5 & placed both hands on the graph, but only used the left hand to feel the peaks at first, came up to the center and then changed the hand and right hand till the center and left hand after during the second feel round. same peak confusion ar the others. \\
\hline
P5 & STG & skewness & no & p2w & L & l[1] & 1 & s1 & placed right hand, but did not use it to discern anything. maybe because they placed their hand their and touched it without feeling the rest of the graph or even consciously feeling it, they thought it was a tail to the right. \\
\hline
P5 & STG & kurtosis & meso & p2w & L/R & l[1],lrci[1],lrco[2] & 4 & s1,s5 & used the same hand strategy \\
\hline
P5 & sonification & modality & multi & l2r & -- & l[1],r[1],l[1] & 3 & -- & took some time between go and pressing the buttons \\
\hline
P5 & sonification & skewness & negative & none & -- & l[1],r[1] & 2 & -- & was very fast \\
\hline
P5 & sonification & kurtosis & platy & none & -- & l[1],r[1] & 2 & -- & pressed the wrong buttons initially, I did not notice \\
\hline
P6 & BDP & modality & multi & none & L & crl[2] & 2 & s1,s5 & did not spend a long time pondering, the right hand was placed on a section of dots constantly \\
\hline
P6 & BDP & kurtosis & platy & none & L/R & l[3] & 3 & s1,s5 & briskly touched everything without fixating on any section \\
\hline
P6 & BDP & skewness & negative & none & L/R & lrco[2] & 2 & s1,s5 & \\
\hline
P6 & sonification & modality & bi & l2r & -- & l[1] & 1 & -- & listened to it only once \\
\hline
P6 & sonification & kurtosis & lepto & l2r & -- & l[1] & 1 & -- & \\
\hline
P6 & sonification & skewness & positive & l2r & -- & l[1] & 1 & -- & \\
\hline
P6 & STG & modality & uni & w2p & L/R & u[1] & 1 & s2,s1 & was very quick and confident, touched the trend line to understand \\
\hline
P6 & STG & kurtosis & meso & w2p & L/R & lrco[1], crl[1] & 2 & s1 & The lrco, first with left hand when going left and right hand when going right, not at the same time \\
\hline
P6 & STG & skewness & no & w2p & L/R & d2u[1], clr[1] & 2 & s1 & \\
\hline
P7 & STG & skewness & no & w2p & L/R & lrco[1], lrci[1], lrco[1], lrci[1], lrco[1], lrci[1], lrco[1], d2u[1], lrco[1] & 9 & s1,s6,s5 & took a lot of time delibrating and retraced his path multiple times. Down to up was only for the peak values \\
\hline
P7 & STG & kurtosis & meso & w2p & L/R & lrci[1], lrco[1], l[1], lrci[1], lrco[1], crl[1] & 6 & s1,s5 & Again checked the peaks multiple times which no one did before \\
\hline
P7 & STG & modality & uni & w2p & L/R & lrci[1], lrco[1], l[1], crl[1], l[1] & 5 & s1,s5 & peak checks for longer time, stayed at the peaks, retracing steps \\
\hline
P7 & sonification & skewness & positive & l2r & -- & l[2],r[1],l[2],r[1], l[1], r[4], l[2], r[1], l[1], r[1], l[1] & 17 & -- & not sure when tone coming from the right \\
\hline
P7 & sonification & kurtosis & lepto & l2r & -- & l[2], r[2], l[1] & 5 & -- & very quick response \\
\hline
P7 & sonification & modality & bi & l2r & -- & l[1], r[1], l[1] & 3 & -- & even more quick \\
\hline
P7 & BDP & skewness & negative & l2r & L & l[1], r[1], l[1], r[1], l[1], r[1], l[1] & 7 & -- & kept the right hand at the edge of the display \\
\hline
P7 & BDP & kurtosis & platy & l2r & L & l[1], r[1], l[1], r[1], l[1], r[1], l[1], r[1] & 8 & -- & same strategy as the one above \\
\hline
P7 & BDP & modality & multi & l2r & L & l[1], r[1], l[1] & 3 & -- & very quick response \\
\label{appendixtable:gesture-analysis}
\end{longtable}